# Using Deep Learning to Predict Beam-Tunable Pareto Optimal Dose Distribution for Intensity Modulated Radiation Therapy

Gyanendra Bohara, Azar Sadeghnejad Barkousaraie, Steve Jiang, Dan Nguyen

Medical Artificial Intelligence and Automation (MAIA) Laboratory, Department of Radiation Oncology, UT Southwestern Medical Center, Dallas TX, USA

E-mail: Dan.Nguyen@UTSouthwestern.edu

We propose to develop deep learning models that can predict Pareto optimal dose distributions by using any given set of beam angles, along with patient anatomy, as input to train the deep neural networks. We implement and compare two deep learning networks that predict with two different beam configuration modalities. We generated Pareto optimal plans for 70 patients with prostate cancer. We used fluence map optimization to generate 500 IMRT plans that sampled the Pareto surface for each patient, for a total of 35,000 plans. We studied and compared two different models, Model I and Model II. Model I directly uses beam angles as a second input to the network as a binary vector. Model II converts the beam angles into beam doses that are conformal to the PTV.

Our deep learning models predicted voxel-level dose distributions that precisely matched the ground truth dose distributions. Quantitatively, Model I prediction error of 0.043 (confirmation), 0.043 (homogeneity), 0.327 (R50), 2.80% (D95), 3.90% (D98), 0.6% (D50), 1.10% (D2) was lower than that of Model II, which obtained 0.076 (confirmation), 0.058 (homogeneity), 0.626 (R50), 7.10% (D95), 6.50% (D98), 8.40% (D50), 6.30% (D2). Treatment planners who use our models will be able to use deep learning to control the tradeoffs between the PTV and OAR weights, as well as the beam number and configurations in real time. Our dose prediction methods provide a stepping stone to building automatic IMRT treatment planning.

## 1. Introduction

Today, it is estimated that about two-thirds of all patients with cancer receive Radiation Therapy as a unique treatment or in combination with more complex treatment procedures.[1] One of the remarkable achievements in External Beam Radiation Therapy (EBRT) is the development of Intensity Modulated Radiation Therapy (IMRT),[2-7] which uses variable beam intensities to treat cancer. IMRT allows the delivery of less dose to the organs at risk (OARs) and more dose to the planning target volume (PTV) than 3D conformal radiation therapy,[8-11] but its treatment planning process is more difficult and time consuming. IMRT treatment planning consists of two iterative processes: first, the planner uses dose-volume constraints and other hyper-parameters to obtain an optimal plan to deliver as much of the prescribed dose to the PTV as possible while minimizing the dose to critical structures. The planner has to iteratively and tediously tune the hyper-parameters in a trial-and-error fashion. Second, the physician reviews the plan and provides further comments and feedback to get the outcome that achieves the best tradeoffs between PTV and OARs.[12,13] These two processes loop until the final plan is approved. It can take from multiple hours to a week to generate an acceptable plan, depending on the treatment site and its complexity.

Several studies have tried to improve the treatment planning process by using mathematical optimization algorithms to account for various aspects. Multicriteria optimization[13-15] focuses on generating multiple plans with tradeoffs between the PTV and OARs on the Pareto surface, which allows the clinician to then choose the plan with their desired tradeoffs. Beam orientation optimization[16-21] focuses on finding a suitable set of beam directions that improves upon manually selected or protocol-based beam orientations. Direct aperture optimization, also called machine parameter optimization,[22-27] focuses on generating deliverable, high quality plans by determining the optimal aperture shapes and their intensities. There are many commercial software packages available based on mathematical optimization algorithms, such as Eclipse™ comprehensive treatment planning (Varian Medical systems, Palo Alto, CA, USA), Pinnacle treatment planning (Philips Radiation Oncology, Fitchburg, WI, USA),[28] and RayPlan treatment planning system (RaySearch Laboratories, Stockholm, Sweden). Still, these systems require manual, tedious tuning of hyper-parameters, such as structure weights, beam geometries, appropriate dose-volume constraints, and tradeoffs between the PTV and OARs. Also, treatment plans generated on these systems differ from planner to planner and from physician to physician based on their work experience and preferences. The customized tedious process, planning variations based on personal experiences and preferences, and the need for strong domain knowledge expertise could lead to suboptimal plans that compromise patient care.[29-31]

A new set of methods, called knowledge-based planning (KBP),[32-37] has been developed to address the shortcomings of mathematical optimization algorithms and improve the quality and efficiency of treatment planning by learning a database of carefully designed past clinical plans. KBP uses machine learning algorithms and is a powerful tool for guiding treatment planners and physicians to achieve high quality plans. RapidPlan[TM] is an example of KBP that was developed by Varian Medical Systems. This system estimates the dose volume histogram (DVH) for the new plan by using patient-specific geometry. Many researchers have reported on RapidPlan's performance and compared it with that of conventional treatment planning, and they have found that, in its current state, RapidPlan is much faster and can generate clinically acceptable plans with higher quality than conventional treatment planning for about half of the cases.[38-43] However, it is not fully automated yet, and for the remaining treatment cases, manual tuning is still necessary to make acceptable plans. In addition, KBP relies heavily on small datasets because datasets have not been integrated between different institutions, so caution should be taken when applying these methods to patients whose geometry falls outside the plan library.[39] Also, before the deep learning era, KBP methods used more traditional machine learning algorithms, and they were limited to predicting DVH or particular dosimetric criteria from user-defined features.[44]

Deep learning has advanced many areas such as image recognition, speech recognition, natural language translation towards automation[45] and has addressed the shortcomings of traditional machine learning by learning its own features from data without the need for human intervention. Likewise, deep learning has the potential to automate the IMRT treatment planning process by removing its dependence on handcrafted features. The development of the fully convolutional network (FCN)[46] allowed for pixel-wise prediction using supervised learning, which opened the door for voxel-wise dose prediction and generation of DVH curves in treatment planning. Recently, many researchers have developed different deep learning models for predicting clinical dose distributions for IMRT and Volumetric Modulated Arc Therapy (VMAT) modalities on different treatment sites such as lung, prostate, and head-and-neck.[47-53] However, all of these models used static beam orientations for their study, thus limiting their uses in the treatment planning workflow to a subset of common treatment plans based on protocol. One approach that uses varying beam angles to predict the clinical dose for lung IMRT patients has been developed recently.[54]

Clinical dose prediction models are often limited to a single dose predicted per patient. This contrasts with Pareto optimal dose prediction models, which can generate multiple plans that have differing tradeoffs between the different critical structures. Previous studies found that deep learning models that use anatomical structures and static beam orientations to predict Pareto optimal dose distributions could generate multiple optimal plans with differing tradeoffs in real time.[44,55] However, Pareto optimal dose predictions for IMRT prostate plans with variable beam numbers and orientations have not yet been studied. In this paper, we present an approach that uses deep learning networks to predict Pareto optimal dose distributions for prostate IMRT plans that involve anatomical structures and varying beam numbers and orientations. Specifically, our contribution to the current literature is the addition of the ability to tune the beam orientations in a deep learning-based, Pareto optimal dose prediction model. Such a model would allow for a treatment planner to quickly explore the beam orientation space, and select a beam arrangement that can even be outside of the typical clinical protocol. We implement and compare two deep learning networks that predict with two different beam configuration modalities: Model I, which uses the direct input of the beam angles in the network as a binary vector, and Model II, which uses the conformal beam dose that corresponds to the beam angles used in Model I. Model II serves here as a state-of-the-art model for comparison; this model is similar to the AB model introduced by Ana et al.,[54] where beam setup information was represented by the cumulative dose distribution for all the beams in the plan computed by using the fluence-convolution broad beam (FCBB)[56] dose calculation method. In our case, we used a simple algorithm to generate a beam conformal to the PTV structure (see section 2.2).

This work will provide treatment planners with the advantage of using deep learning to control the tradeoffs between the PTV and OAR weights, as well as the beam number and configurations, in real time.

## 2. Materials and Methods

For this study, we generated Pareto optimal plans for 70 patients with prostate cancer. We used fluence map optimization to generate 500 IMRT plans that sampled the Pareto surface for each patient, for a total of 35,000 plans. More details about generating the Pareto optimal dose distribution dataset are presented in section 2.1. The deep learning models used for predicting Pareto optimal dose distributions are described in section 2.3. We studied and compared two different models. Although they both used the same anatomical structures—which included the planning target volume (PTV), organs at risk (OARs), and body—these models were designed with two different methods for representing the beam angles. For Model I, we directly used beam angles as a binary vector for the second input to the network. For Model II, we converted the beam angles into beam doses that were conformal to the PTV. More details about generating the conformal beam doses are provided in section 2.2. We divided the 70 patients into 54 training, 6 validation, and 10 testing patients, yielding 27,000 training, 3,000 validation, and 5,000 testing plans. Detailed explanations of the model training, validation, and testing are presented in section 2.4.

## 2.1 Pareto Optimal Plans

The Pareto optimal solutions for 70 patients with prostate cancer were generated by minimizing the objective function defined from Equations 1-3. These resulting dose distributions for training are calculated after the fluence map optimization step, but before any machine parameter sequencing is performed. The parameters involved were anatomical structures and 10 different sets of 1 to 10 random beam angles. Anatomical structures included the planning target volume (PTV) and the organs at risk (OARs): body, bladder, rectum, left femoral head, and right femoral head. Shell and skin structures were also included in the plan as tuning structures. Pareto optimal solutions represent the various tradeoffs between tumor coverage and normal tissue sparing. This is associated with a multicriteria objective, which can be written as

$$\begin{aligned} \underset{x_\theta}{minimize} \quad & \{f_{PTV}(x_\theta), f_{OAR_1}(x_\theta), f_{OAR_2}(x_\theta), \dots, f_{OAR_n}(x_\theta)\} \\ subject\ to \quad & x_\theta \geq 0, \quad \theta \epsilon A \end{aligned} \quad (1)$$

where $f_s$ is the objective function and $A = [\theta_1, \dots \theta_{10}]$ is the collection of all 10 sets of beam angles. For example, $\theta_1 = [10]$, $\theta_2 = [24,38], \dots, \theta_{10} = [20,26,30,38,46,6,56,98,64,120]$ are the randomly generated angles, and $x_\theta$ refers to the fluence map intensities to be optimized. Here, we use the $\ell_2$ -norm to formulate the objective,

$$f_s(x_\theta) = \frac{1}{2} \left\| d_{\theta,s} x_\theta - p_s \right\|_2^2 \quad (2)$$

where $d_{\theta,s}$ is the dose influence matrix for the $\theta_{th}$ beam and the $s_{th}$ structure. The dose influence matrices were determined using 1 to 10 random coplanar beams where the beamlet size was 2.5 mm$^2$ at a 100 cm isocenter. $p_s$ is the desired dose for a given structure, assigned as the prescription dose if $s$ is the PTV, and otherwise 0. The dose influence calculation was performed using the Analytical Anisotropic Algorithm (AAA) provided by the Eclipse treatment planning system, using the built-in application programming interface (Varian Medical systems, Palo Alto, CA, USA). Now, we can reformulate the multicriteria optimization[14,57,58] as a single-objective, convex optimization problem:

$$\underset{x_\theta}{minimize} \qquad \sum_{s \in S} w_s^2 f_s(x_\theta) \qquad (3)$$

$$subject\ to \qquad x_\theta \geq 0, \quad \theta \epsilon A$$

where $w_s$ are the user-defined tradeoff weights for each structure. Different Pareto optimal plans can be generated by varying the $w_s$ to different values. We generated many pseudo-random plans by assigning random weights, as described below, to the organs at risk by using an in-house GPU-based proximal-class first-order primal-dual algorithm, Chambolle-Pock.[59] While there are some uses of the Chambolle-Pock algorithm used in radiation therapy[60-64], any convex solver will theoretically arrive at the same solution as Chambolle-Pock for solving the optimization problem. The weight generation structure fell into one of three categories, as shown in Table 1.

| Category | Description |
| --- | --- |
| Low weights | $w_s = rand(0,0.1) \ \forall s \in OAR$ |
| Extra low weights | $w_s = rand(0,0.05) \ \forall s \in OAR$ |
| Controlled weights | $w_{bladder} = rand(0,0.2)$ |
| | $w_{rectum} = rand(0,0.2)$ |
| | $w_{lt\ fem\ head} = rand(0,0.1)$ |
| | $w_{rt\ fem\ head} = rand(0,0.1)$ |
| | $w_{shell} = rand(0,0.1)$ |
| | $w_{skin} = rand(0,0.3)$ |

Table 1: Weight generation categories for the organs at risk. The function $rand(L_B, U_B)$ creates a uniform random number between a lower bound ($L_B$) and an upper bound ($U_B$). In all categories, the PTV weights were assigned 1.

For each patient, we created 500 plans spanning the low, extra low, and controlled weights categories. These bounds for the controlled weights were chosen through a trial-and-error method so that the plan generated would fall within clinically relevant limits, even though it is not necessarily acceptable by a physician. A total of 35,000 IMRT plans were created, each as 96 x 96 x 32 dimension arrays with a voxel size of 5 mm$^3$ that sampled the Pareto surface.  Table 2 shows the distribution of Pareto optimal plans generated in each weight category per beam set and its assignment as training, validation and testing for the study.

| Weights | Training Plans per beam set | Validation Plans per beam set | Testing Plans per beam set |
|---|---|---|---|
| Low | 1080 | 120 | 200 |
| Extra Low | 540 | 60 | 100 |

| Controlled | 1080 | 120 | 200 |

Table 2: Distribution of Pareto plans, 1) 40% of plans were assigned with low weights, 2) 20% of plans were assigned extra low weights, 3)40% of plans were assigned with controlled weights

## 2. Conformal Beam Dose Data

Conformal beam dose describes the high dose volume that is shaped to closely conform to the desired PTV structure. There are numerous ways to make a broad beam conform to the PTV.[56,65] In this study, we used a simple method to generate a beam that is conformal to the PTV structure. For each beam we first selected a square field of $20 \times 20$ beamlets, with dimensions of $2.5 \, mm \times 2.5 \, mm$ per beamlet. This the same beamlet data mentioned in section 2.1 that was generated using AAA dose calculation algorithm. We then scale this beam such that its mean dose contribution to the PTV is equal to the prescription dose. Beamlets with an integral dose contribution of less than the threshold of 1% of the prescription dose to the PTV are removed. The remaining beamlets then create a conformal beam around the PTV.

For each patient, 500 plans were generated using all 10 sets of beam angles. Representative images of conformal doses are shown in Figure 1.

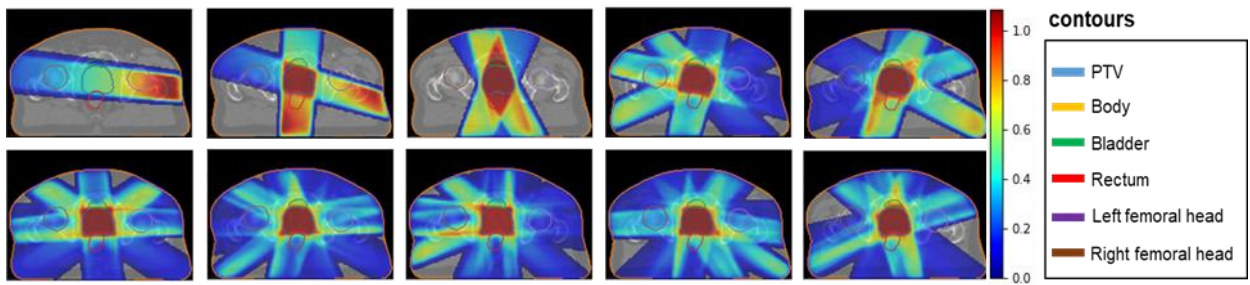

**Figure 1:** Conformal dose corresponding to different beam angles (1-10).

## 2. 3. Network Architecture

The network architecture used in this study is depicted in Figure 2. The dose prediction models used a U-Net style architecture.[66] We used group normalization[67] instead of batch normalization. This network consists of three major parts: downsampling, bottom, and upsampling. The rectified linear unit (ReLU), group normalization and dropout were applied immediately after every convolution operation in the hidden layers. For the sake of clarity, the following paragraphs will assume these operations are included when "convolution" is mentioned. More details on ReLU, group normalization, and dropout are mentioned later, after the main network architecture description. With the exception of the strided convolution, all other convolutions are zero-padded before the convolution, to maintain the same data shape.

The downsampling part of the U-Net is constructed to contain a five-level hierarchy. We chose five levels with four $2 \times 2 \times 2$ downsampling operations to halve the feature size four times, reducing the data from $96 \times 96 \times 32$ voxels to $6 \times 6 \times 2$ voxels. In each level, two $3 \times 3 \times 3$ convolution operations took place, followed by one downsampling operation. The downsampling operation is a concatenation of two other operations: 1) $2 \times 2 \times 2$ max pooling and 2) $2 \times 2 \times 2$ strided convolution with a $2 \times 2 \times 2$ kernel. In this process, feature channels were doubled in each level.

The bottom part is between the downsampling and upsampling parts of the network. This part takes the last downsampled feature map and performs two $3 \times 3 \times 3$ convolutions. In addition, for one of the models to be evaluated (Section 2.3.1), beam angle information is added as a binary vector that is then processed through a fully connected network. This data is reshaped and concatenated onto the bottom level, prior to the two convolutions.

The upsampling part of the network also consists of five levels, just like the downsampling part. The purpose of this part is to combine the features and spatial information through a sequence of upsampling 3x3x3 and convolution operations and to concatenate high resolution features from the downsampling part. This part consists of 4 upsampling layers, in addition to the final convolution output layer. In this process, feature channels are reduced by half, but feature size is doubled after each convolution to maintain symmetry and original feature size.

All activation functions in the hidden layers are rectified linear units (ReLU), but the final activation function is a linear activation function. Group normalization was used in all hidden layers after the convolution and ReLU operations, which normalizes the weights by grouping feature channels of 32, 64, 128, 256, and 512 into 1, 2, 4, 8, and 16 groups, respectively, of 32 channels each (Fig. 2); this allows faster convergence. The dropout scheme, from a previous paper[47], described in Table 3 was applied after each group normalization.

| U-net Hierarchy Level | Groups | Dropout Rate |
| --- | --- | --- |
| 1 | 1 | 0.125 |
| 2 | 2 | 0.148 |
| 3 | 4 | 0.176 |
| 4 | 8 | 0.210 |
| 5 | 16 | 0.250 |

Table 3: Dropout rate scheme used in the networks.

Models used in this investigation are of two types.

**2. 3.1. Model I**

The models used in this study are depicted in Figure 2. Model I's architecture is exactly what is shown in Figure 2. Model 1 takes in 3 channels: a PTV channel, a body channel, and an OARs channel. Instead of just binary masks as input, we multiply the masks with their corresponding weights, $w_s$, from Equation 3, such that a voxel is defined as $w_s$ if a voxel is defined inside a structure and 0 otherwise. The PTV and body channels have just their respective data, while the OARs channel contains the bladder, rectum, femoral heads, and tuning structures information. Mentioned in Section 2.1, the data was resized to 5 mm³ voxels. To maintain uniform data shape for model training, all patient data was filled into a $96 \times 96 \times 32$ array. The body segmentation covers CT slices. The input data for Model 1 ranges from 0 to 1 since the structure weights used for optimization, $w_s$, were also defined from 0 to 1. As the second input, randomly generated beam angles as a Boolean array of 180 elements—representing angles with 2 degrees separation—are concatenated in the bottom part of the network. In the Boolean array, the current selected beam angles are considered as ones, and all other unselected beam angles are considered as zeros. The Boolean array of 180 elements was the input of a fully connected layer with output of 2304 elements. After that a reshaped operation was applied to change the 2304 length data to $(6,6,2,n)$ where n was 32 (i.e., $2304 = 6 \times 6 \times 2 \times 32$). The reshaped beam angles were matched to the downsampling feature size, allowing for them to be concatenated together along the channels axis, for further processing in the network.

## 2. 3.2 Model II

Model II's architecture is the same as shown in Figure 2 except without the beam angle binary vector input in the bottom. Model II consists of a single four-channel input: the first three channels are the same as Model I, and the last channel is the conformal dose information for a set of selected beam angles (see section 2.2). The size of each input channels is also $96 \times 96 \times 32$. Analogous to Model I, the anatomical inputs range from 0 to 1. For the additional conformal dose channel, the dose was divided by its maximum dose, to also range from 0 to 1.

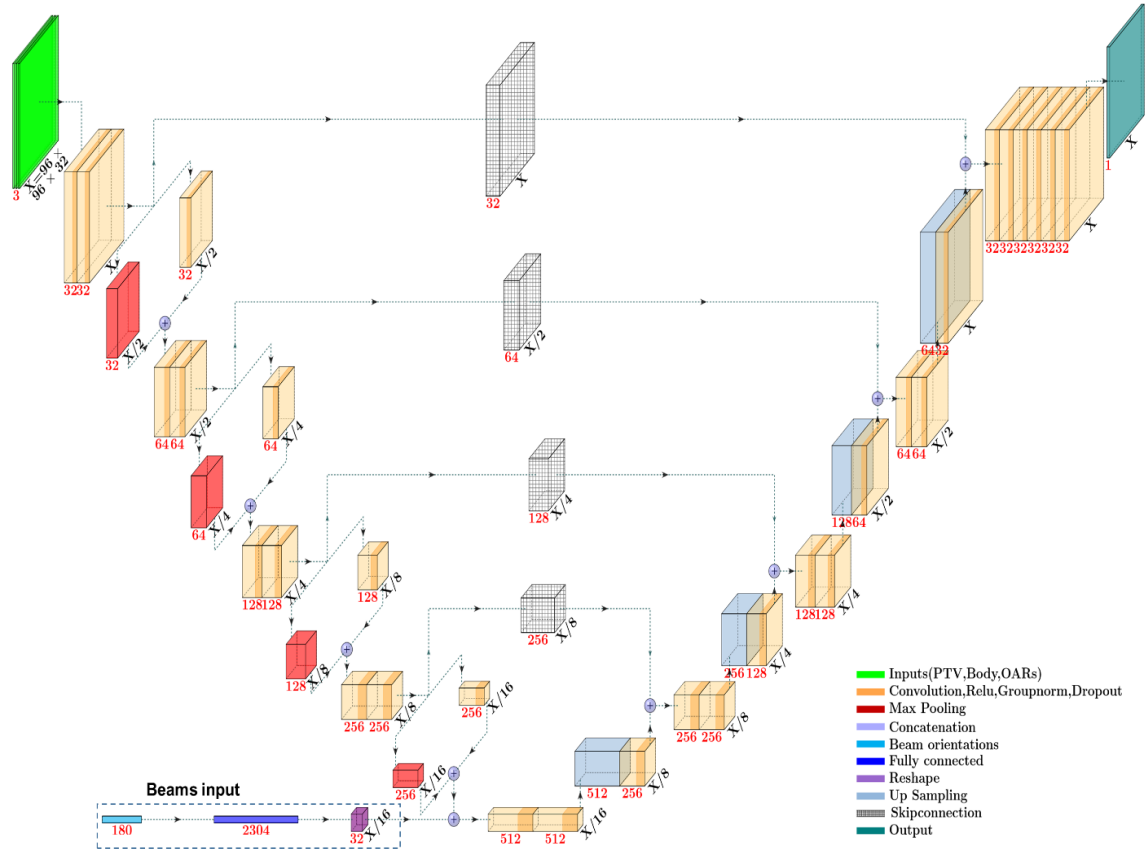

**Figure 2:** Deep learning models used in the study.

## 2. 4. Model Training, Validation, and Testing

For each model, we divided the 70 patients into 54 training, 6 validation, and 10 testing patients, thus yielding 27,000 training, 3,000 validation, and 5,000 testing plans. We implemented a maximum dropout rate of 0.25 to regularize the network and, thus, avoid overfitting. We applied these dropout rates after group normalization so that the highest group, 16, got the dropout rate of 0.25 and the lowest group, 1, got 0.125. The dropout rate scheme is presented in Table 3. During training, we used a batch size of 1, which was due to memory constraints.

The mean square loss (MSE)

$$MSE = \frac{1}{n}\sum_{v}^{m}(D_{test}^{v} - D_{pred}^{v})^2 \qquad (4)$$

was taken as a loss function, where $v$ refers to the voxel index and $m$ represents the total number of voxels. We used the Adam optimizer[68] with a default learning rate of 0.01 to optimize the network's performance. All training was performed on an NVIDIA Quadro P6000 GPU with 24 GB RAM. The models were trained with an early stopping scheme,[69] which is a regularization method that prevents overfitting. This scheme stops the network training when the model is no longer improving the validation loss after a set number of iterations, then it saves the model with the lowest validation loss. For our study, we trained the model for an additional 40,000 iterations after finding the best performing model and terminated the training process if no further improving was observed. The validation loss was checked every 100 iterations. The best models with the lowest total validation loss were used to work out the test data after the completion of training.

We evaluated the models' performance by comparing their predicted dose distributions with the Pareto optimal dose distribution (ground truth) in terms of DVH plots and evaluation metrics, such as PTV $D_{98}$, $D_{95}$, $D_{50}$, $D_{2}$, Paddick Conformation Number,[70] R50 and Homogeneity index and the structure max and mean doses ($D_{max}$ and $D_{mean}$). Readers should refer to the literature for more details about these evaluation metrics.[47,54,71] $D_{max}$ is considered as the dose delivered to 2% of the structure volume, as recommended by the ICRU report.[72] We also compared the predicted dose distributions with the ground truth dose distribution using dose map differences in the clinically relevant PTV and OARs regions.

In addition we have shown an example of beam tuning in the treatment plan with 9 fields Protocol Based IMRT setup.

## 3. Results

The instance of Model I with the lowest validation loss was found at 356,500 iterations, which took about 175 hours to obtain with the early stopping scheme. The instance of Model II with the lowest validation loss was found at 132,300 iterations, which took about 65 hours to obtain. Further iterations after the lowest validation model did not improve the result. After training, the prediction time of each model is less than 1 second. The loss versus iterations evaluated for the training (blue line) and validation sets (red line) from both models are presented in Figure 3. Observing these loss trends gives us an idea of the models' performance. We observed the decreasing trend of losses during training and validation until the best model was achieved. Each model achieved similar MSE losses, with training losses at $1.007 \times 10^{-4}$ (Model I) and $1.463 \times 10^{-4}$ (Model II) and validation losses at $1.251 \times 10^{-4}$ (Model I) and $1.469 \times 10^{-4}$ (Model II).

Figure 4 shows the colored dose wash distributions in the PTV and in the organs at risk. These distributions are overlaid with the original CT slices. Each row of dose distributions in Figure 4 represents a different treatment plan. Visually, the dose distributions that are predicted from Models I and II are similar to that of the ground truth. Particularly, they are better matched in the high dose region surrounding the PTV, and have less accuracy in the lower dose regions.

Figure 5 shows the dose map differences between the predicted dose distributions and the ground truth dose distributions. This treatment plan is exactly same plan as mentioned in the bottom row of Figure 4. Visually, the color intensity indicates that the dose differences predicted from Model I are better matched with the ground truth than the predicted dose distribution from Model II.

DVHs obtained from dose predictions for the two representative plans are presented in Figure 6. The DVH curves obtained from Model I corresponding to PTV are better matched with the ground truth than the DVH curves obtained from Model II corresponding to PTV, especially in the shoulder of the PTV DVH. The DVH curves corresponding to other critical structures except without body show the fluctuations in the prediction accuracy from both models. Visually, from the test cases in Figure 6, it is not fully clear whether one model outperforms the other in dose prediction accuracy to the OARs.

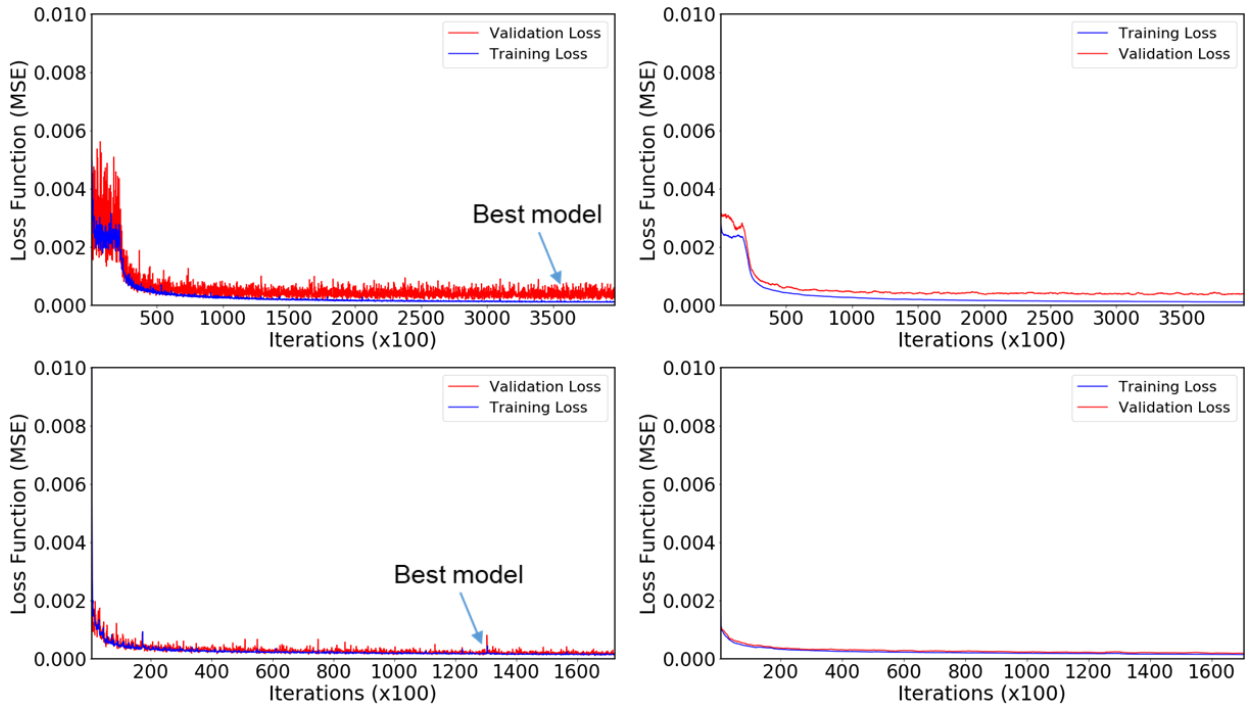

**Figure 3:** Training vs. Validation loss as a function of iterations for both models. Top row plots are for Model I, and bottom row plots are for Model II. Left column plots represent the actual training and validation loss, and the right column plots represent the smooth training and validation losses obtained by using the moving average method.

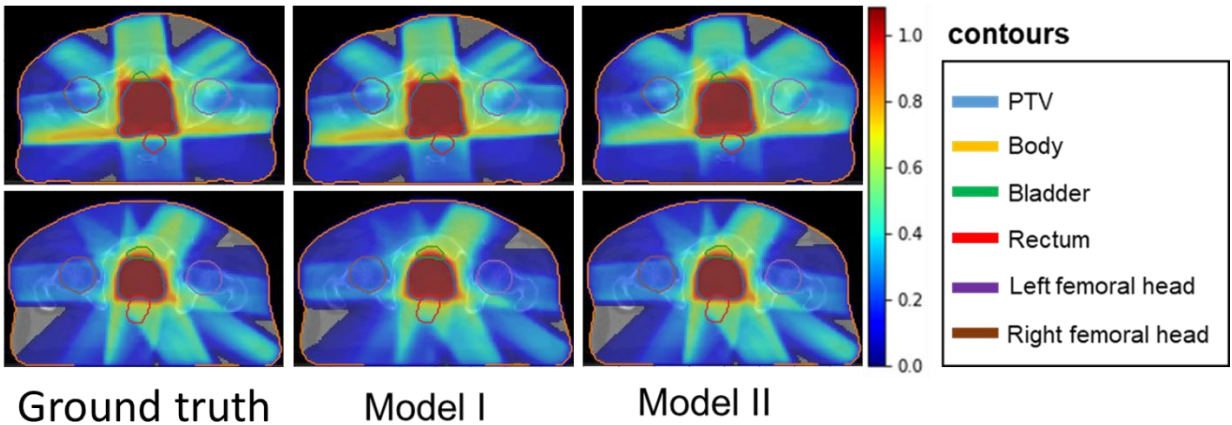

**Figure 4.** Ground truth dose distribution vs. dose distribution predicted by Models I and II. The figures in the first column represents the ground truth dose distributions for two different treatment plans. The second column and third column distributions were predicted by Model I and Model II, respectively.

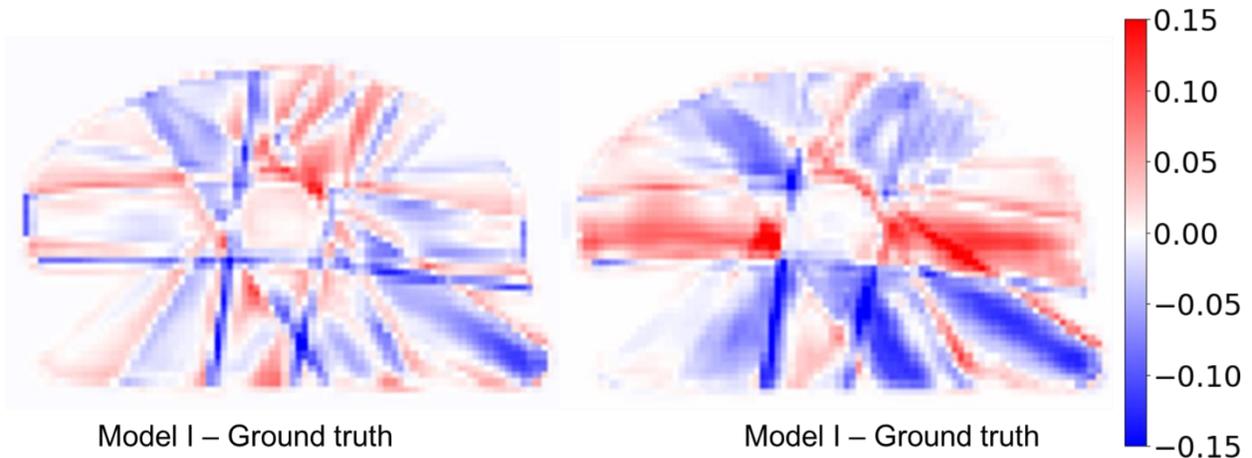

**Figure 5.** Dose map differences between the Model I and Model II predicted dose distributions and the ground truth dose distribution. These differences are obtained taking the differences between Model I and Ground truth, and Model II and Ground truth as shown in second row of Figure 4.

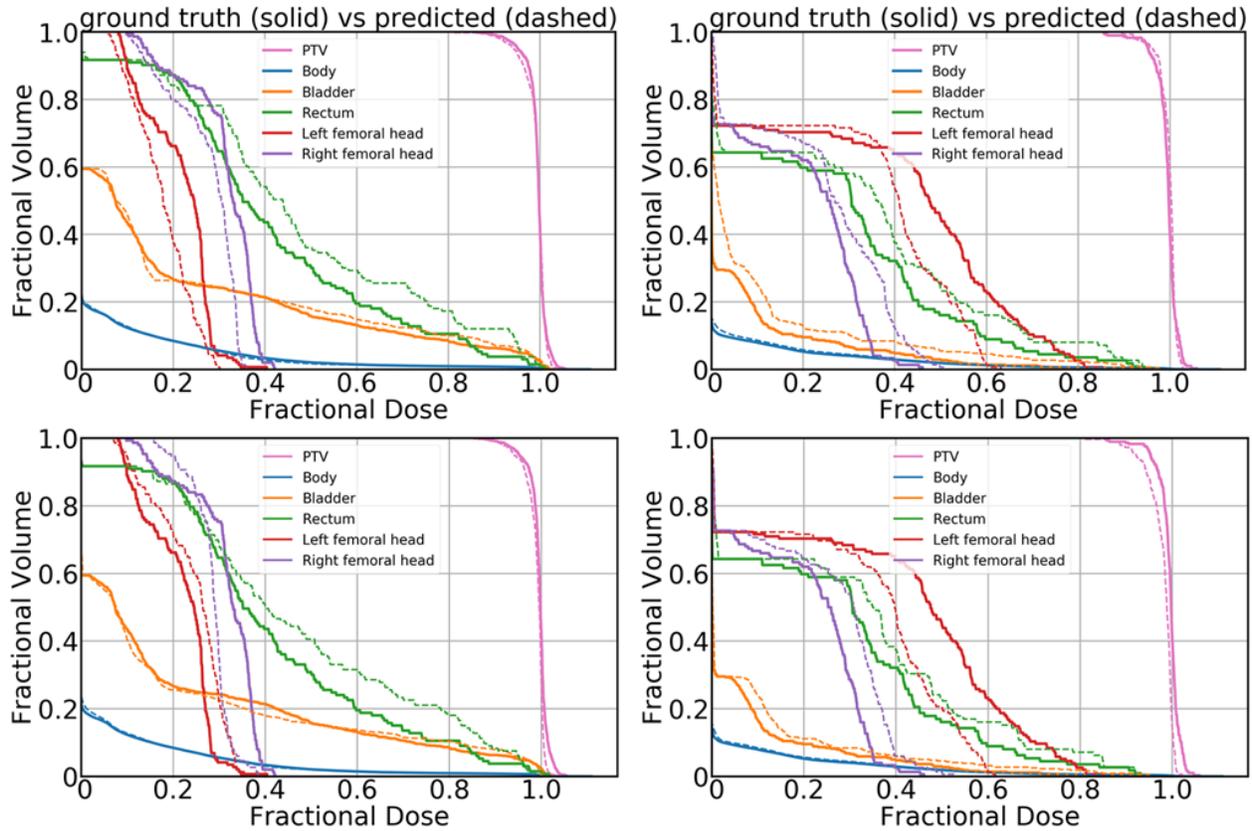

**Figure 6.** DVH plots obtained from Model I (top row) and Model II (bottom row). The top row plots were obtained from Model I, and the bottom row plots were obtained from Model II. The solid lines correspond to the ground truth dose, and the dashed lines correspond to the predicted dose.

Dose evaluation metrics are calculated for all 5000 plans from each testing and predicting dataset. Also, these values are the mean values and deviation from the mean values from all the plans containing 1-10 beam orientations. The metrics presented in the Table 4 represent the dose coverage in the PTV, conformity, dose spillage and homogeneity. For the PTV D98, D95, D50 and D2, the highest mean difference we obtained from Model I was less than 4% and the highest mean difference we obtained from Model II was less than 9%, of the prescription dose. Similarly, for other parameters such as Paddick Confirmation number, dose spillage (R50) and PTV homogeneity, predicted mean value differences are less for Model I in comparison to that of Model II. Overall, the predicted mean differences obtained from Model I are less than that of predicted mean differences obtained from Model II.

The absolute mean dose ($D_{mean}$) and max dose ($D_{max}$) values reported in Table 5 give us an idea of how the dose distributed over the voxels of PTV and other critical structures. For $D_{mean}$ values, the highest mean difference we obtained from Model I was less than 2% and the highest mean difference we obtained from Model II was less than 6%, of the prescription dose. Similarly, for $D_{max}$ values, the highest mean difference we obtained from Model I was less than 5% and the highest mean difference we obtained from Model II was less than 12%, of the prescription dose. In all cases, the prediction errors obtained from Model I are less than that of Model II.

A paired t-test was used to determine if there is a statically significant difference between the performance of Model I and Model II, with respect to how accurately they predicted the ground truth Pareto optimal dose. The largest p-value that we obtained from a two tailed paired t-test was $6.82 \times 10^{-63}$.

| | Pareto optimal dose | Model I Predicted dose | Model II Predicted dose | Model I \|Predicted – Ground truth\| | Model II \|Predicted – Ground truth\| |
|---|---|---|---|---|---|
| | Mean ± SD | Mean ± SD | Mean ± SD | Mean ± SD | Mean ± SD |
| PTV D98 | 0.75±0.08 | 0.75±0.09 | 0.69±0.09 | 0.04±0.03 | 0.07±0.05 |
| PTV D95 | 0.82±0.06 | 0.83±0.06 | 0.76±0.06 | 0.03± 0.03 | 0.07±0.05 |
| PTV D50 | 0.99±0.01 | 0.99±0.01 | 0.90±0.03 | 0.01±0.01 | 0.08±0.02 |
| PTV D2 | 1.02±0.04 | 1.03±0.04 | 0.96±0.02 | 0.01±0.01 | 0.06±0.05 |
| Paddick Confirmation number | 0.63±0.25 | 0.66±0.26 | 0.59±0.21 | 0.04±0.04 | 0.08±0.07 |
| R50 | 5.2±1.3 | 5.0±1.4 | 5.4±1.2 | 0.33±0.23 | 0.63±0.55 |
| PTV Homogeneity | 0.29±0.11 | 0.29±0.12 | 0.30±0.10 | 0.04±0.04 | 0.06±0.05 |

**Table 4:** Means and standard deviations (Mean ± SD) for clinical DVH metrics of ground truth (Pareto optimal) dose, predicted dose, absolute difference between the predicted dose and the ground truth, conformation, high dose spillage (R50) and homogeneity obtained from Model I and Model II.

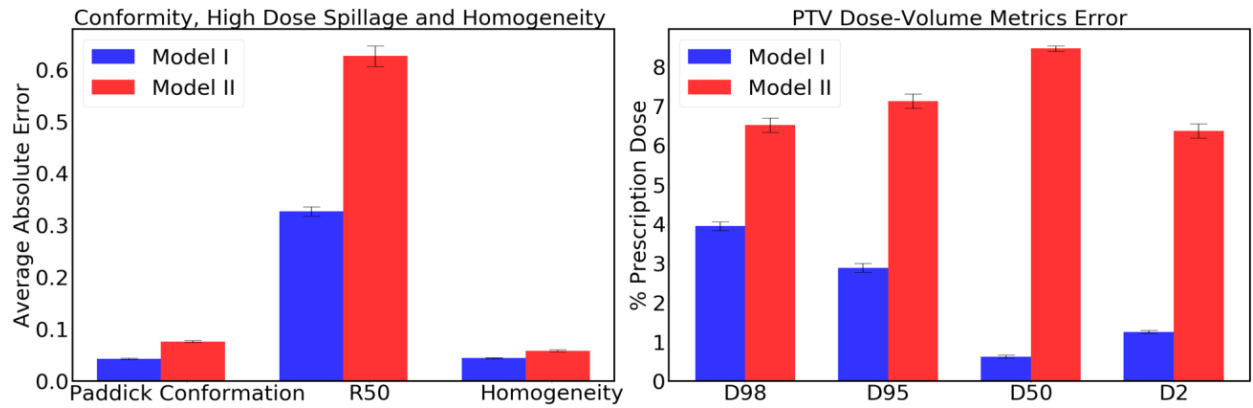

**Figure 7:** Prediction errors obtained from Models I and II for conformation, high dose spillage (R50), homogeneity, and PTV dose coverage on the test data. Error bar represents the 99% confidence interval ($\bar{x} \pm 2.576 * \frac{\sigma}{\sqrt{n}}$), where $\bar{x}$ and $\sigma$ are mean and standard deviation, respectively.

Figure 7 was obtained based on the information presented in Table 4. This shows the errors for several clinical metrics evaluated from the predicted dose distributions, as compared to the metrics of the Pareto optimal dose distributions. Model I's prediction error of 0.043 (confirmation), 0.043 (homogeneity), 0.327 (R50), 2.80% (D95), 3.90% (D98), 0.6% (D50), 1.10% (D2) was lower than that of Model II, from which we obtained 0.076 (confirmation), 0.058 (homogeneity), 0.626 (R50), 7.10% (D95), 6.50% (D98), 8.40% (D50), 6.30% (D2). In terms of these dosimetric criteria, Model I performed better than Model II.

|  |  | Pareto optimal dose | Model I Predicted dose | Model II Predicted dose | Model I \|Predicted − Ground truth\| | Model II \|Predicted − Ground truth\| |
|---|---|---|---|---|---|---|
|  |  | Mean ± SD | Mean ± SD | Mean ± SD | Mean ± SD | Mean ± SD |
| $D_{max}$ | PTV | 1.02±0.04 | 1.03±0.04 | 0.96±0.02 | 0.01±0.01 | 0.06±0.05 |
|  | Body | 0.54±0.01 | 0.53±0.15 | 0.52±0.12 | 0.02± 0.02 | 0.06±0.05 |
|  | Bladder | 0.96±0.08 | 0.97±0.08 | 0.88±0.06 | 0.01±0.01 | 0.09±0.06 |
|  | Rectum | 1.00±0.05 | 1.00±0.05 | 0.93±0.04 | 0.01±0.01 | 0.07±0.05 |
|  | Left fem | 0.47±0.30 | 0.46±0.30 | 0.47±0.25 | 0.05±0.06 | 0.12±0.11 |
|  | Right fem | 0.35±0.28 | 0.33±0.27 | 0.32±0.23 | 0.03±0.04 | 0.07±0.08 |
| $D_{mean}$ | PTV | 0.87±0.08 | 0.87±0.08 | 0.82±0.07 | 0.01±0.01 | 0.06±0.03 |
|  | Body | 0.03±0.01 | 0.03±0.01 | 0.03±0.01 | 0.00±0.00 | 0.00±0.00 |
|  | Bladder | 0.22±0.11 | 0.22±0.10 | 0.20±0.10 | 0.01±0.01 | 0.02±0.02 |
|  | Rectum | 0.51±0.14 | 0.51±0.14 | 0.48±0.12 | 0.02±0.02 | 0.05±0.04 |
|  | Left fem | 0.18±0.15 | 0.18±0.15 | 0.19±0.14 | 0.02±0.03 | 0.05±0.05 |
|  | Right fem | 0.16±0.16 | 0.16±0.15 | 0.16±0.15 | 0.02±0.02 | 0.03±0.04 |

Table 5: Mean and standard deviation (Mean ± SD) of maximum and mean values of the Pareto optimal dose distribution, the predicted dose distribution, and the absolute difference between the predicted dose distribution and the ground truth received on the PTV and other critical structures.

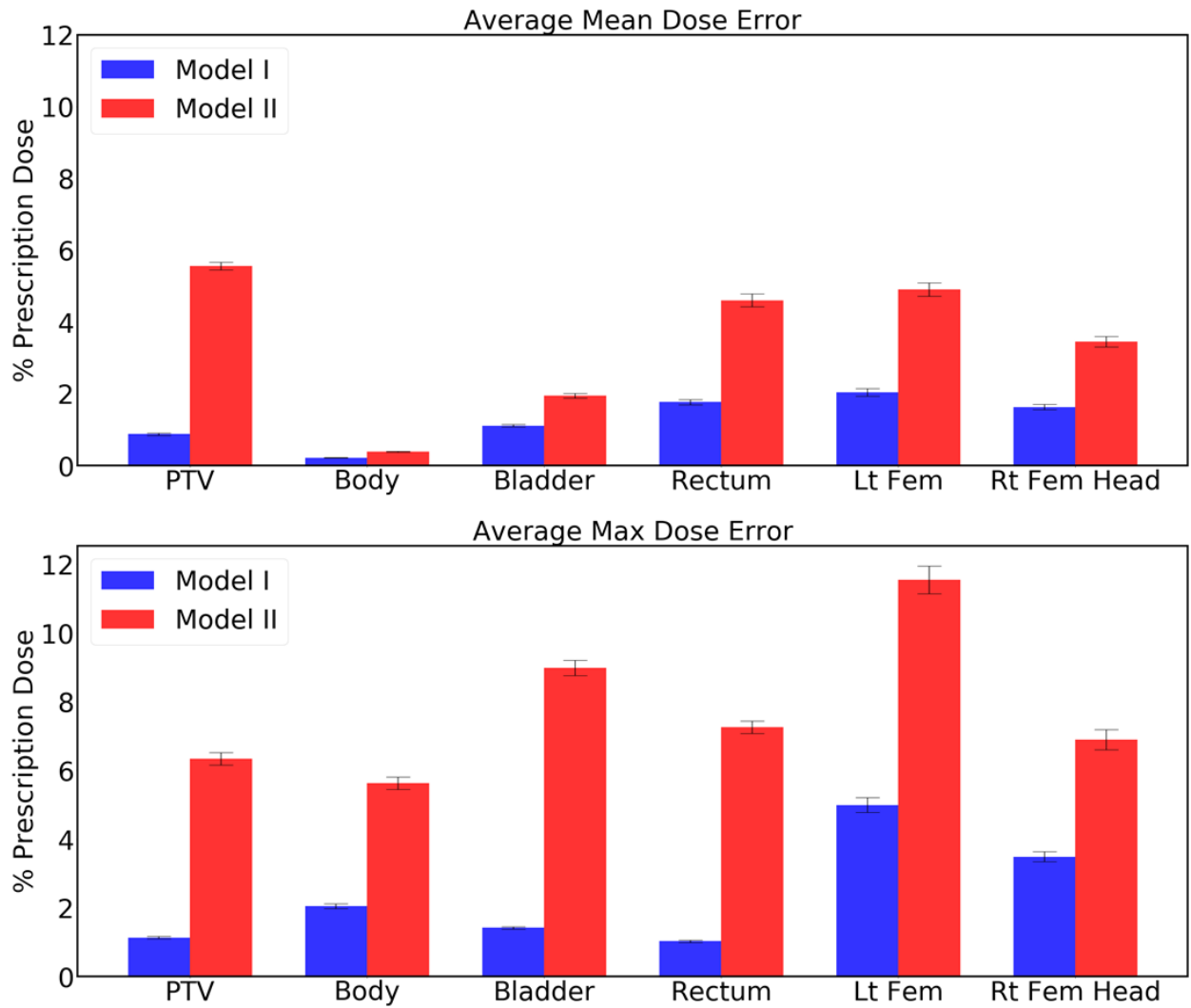

**Figure 8:** Average error in the mean dose (top plot) and the max dose (bottom plot) for the PTV and the organs at risk. Error bar represents the 99% confidence interval ($\bar{x} \pm 2.576 * \frac{\sigma}{\sqrt{n}}$), where $\bar{x}$ and $\sigma$ are mean and standard deviation, respectively.

Figure 8 shows the errors for mean dose and max dose evaluated from the predicted dose distributions, as compared to the metrics of the Pareto optimal dose distributions. Model I had low prediction errors of average mean dose ($D_{mean}$) 0.871% (PTV), 0.214% (Body), 1.10% (Bladder), 1.76% (Rectum), 2.03% (Left Femoral Head), and 1.62% (Right Femoral Head), and average max dose ($D_{max}$) errors of 1.13% (PTV), 2.04% (Body), 1.41% (Bladder), 1.02% (Rectum), 5% (Left Femoral Head), and 3.48% (Right Femoral Head). Model II had relatively high prediction errors of $D_{mean}$ 5.56% (PTV), 0.382% (Body), 1.94% (Bladder), 4.60% (Rectum), 4.90% (Left Femoral Head), and 3.45% (Right Femoral Head), and average $D_{max}$ errors of 6.33% (PTV), 5.62% (Body), 8.98% (Bladder), 7.25% (Rectum), 11.54% (Left Femoral Head), and 6.89% (Right Femoral Head).

As with the dosimetric criteria shown in Figure 7, Model I outperformed Model II for both the mean dose and the max dose errors on PTV, Body, Bladder, Rectum, Left Femoral Head and Right Femoral Head.

| Voxel Wise Dose Difference | Model I | Model II |
|---|---|---|
| D-mean | \|Predicted – Ground truth\| (Mean± std) | \|Predicted – Ground truth\| (Mean± std) |
| Body | 0.0061±0.0018 | 0.0090±0.0029 |
| 10% volume isodose | 0.0049±0.0011 | 0.0088±0.0022 |

Table 6: Mean and standard deviation (Mean ± SD) of the Body and 10 % volume isodose of the absolute voxel based dose differences between predicted and ground truth dose distribution.

The absolute mean dose ($D_{mean}$) values reported in Table 6 give us an idea of how the dose distributed over the voxels of body and 10% volume isodose. For $D_{mean}$ values, the highest mean differences we obtained from Model I and Model II were less than 1%, of the prescription dose. In all cases, the prediction errors obtained from Model I are less than that of Model II.

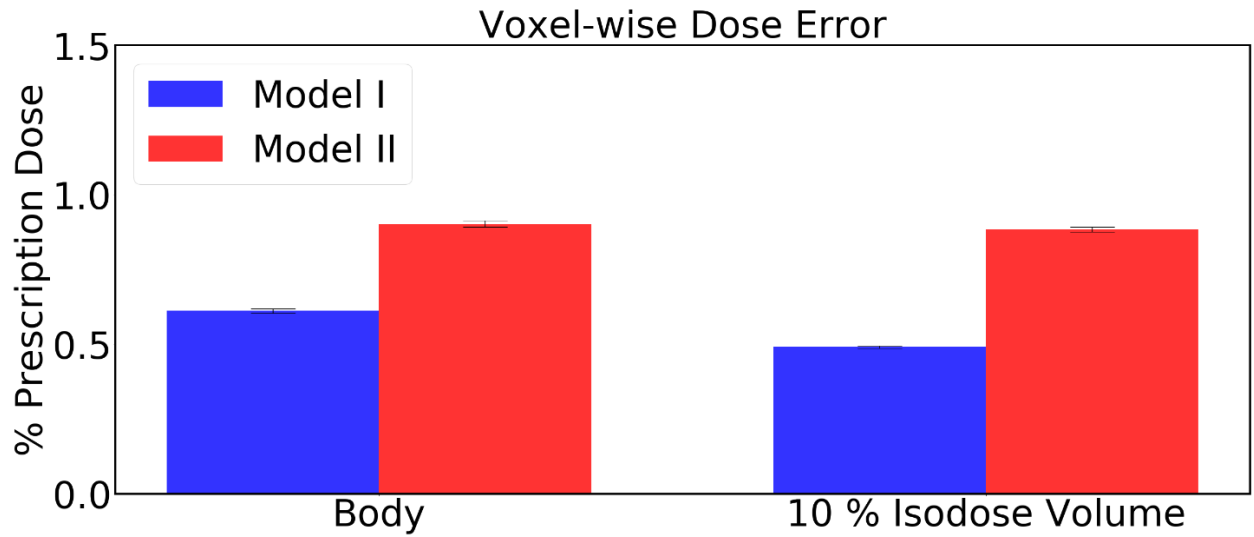

**Figure 9:** Average voxel wise dose error for the Body and the 10% isodose volume. Error bar represents the 99% confidence interval ($\bar{x} \pm 2.576 * \frac{\sigma}{\sqrt{n}}$), where $\bar{x}$ and $\sigma$ are mean and standard deviation, respectively.

Figure 9 shows the Average voxel wise dose errors for the Body and the 10% isodose volume obtained from the predicted dose distributions, as compared to the metrics of the Pareto optimal dose distributions. Model I had low prediction errors of average mean dose ($D_{mean}$) 0.61% (Body), 0.49% (10% isodose volume). Model II had relatively high prediction errors of $D_{mean}$ 0.9% (Body), 0.8% (10% isodose volume). In terms of these dosimetric criteria as well, Model I performed better than Model II.

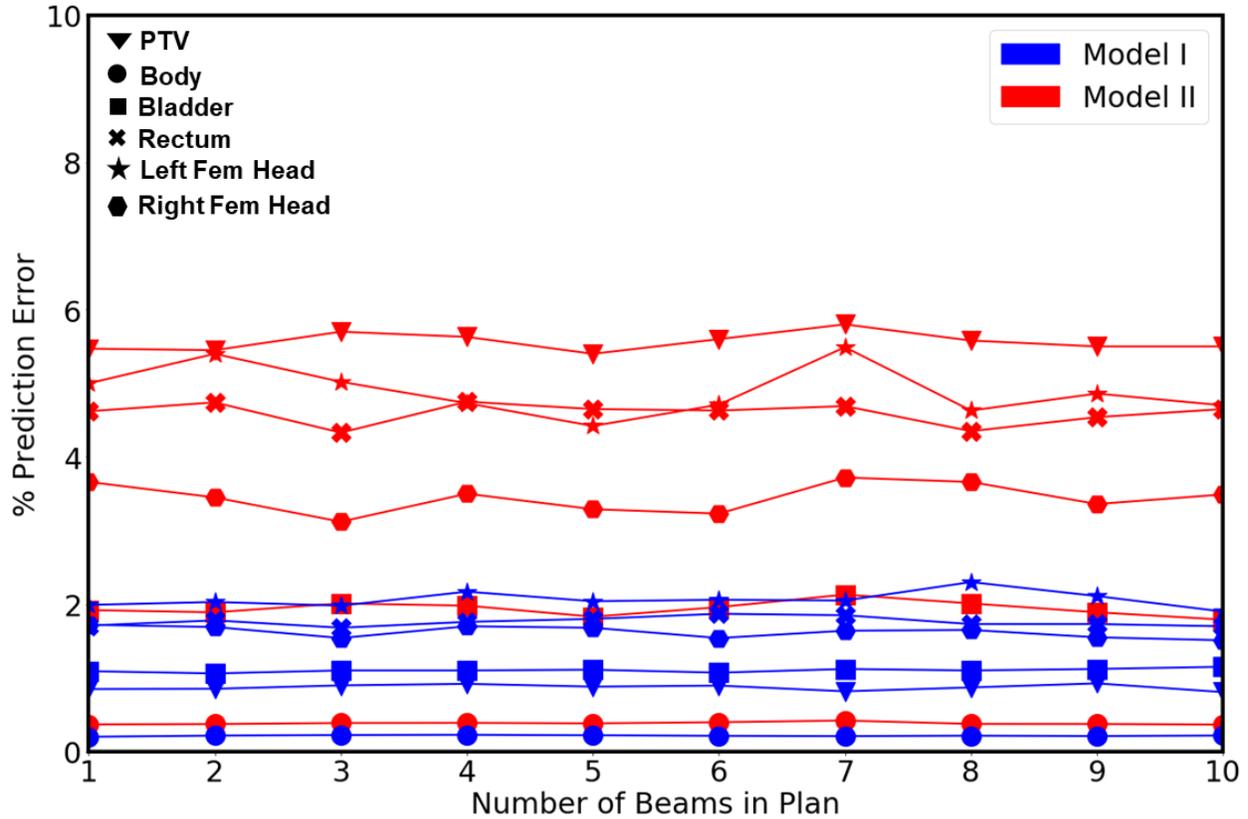

**Figure 10:** Average error in the mean dose for the PTV and the organs at risk obtained from Model I and II corresponding to number of beams in plan geometry (1-10).

Figure 10 shows the errors for mean dose evaluated from the predicted dose distributions, as compared to the Pareto optimal dose distributions corresponding to each beam numbers in plan. It can be seen that that the prediction error is relatively even regardless of the number of beams in the plan. These values corresponding to each beam geometry set are agree with that of average for all 10 sets of beam geometries shown in Figure 8. Model I had low prediction errors of average mean dose ($D_{mean}$) 0.871% (PTV), 0.214% (Body), 1.10% (Bladder), 1.76% (Rectum), 2.03% (Left Femoral Head), and 1.62% (Right Femoral Head). Model II had relatively high prediction errors of $D_{mean}$ corresponding to each beam numbers in plan, with less than 0.40% (Body), 2.00% (Bladder) which are uniform with average prediction errors for all sets of beam geometries reported in Figure 8. For PTV and the rest of the OARs, the prediction errors corresponding to each beam numbers in plan are within 6.00% (PTV), (5.00% (Rectum), 6.00% (Left Femoral Head), and 4.00% (Right Femoral Head) with maximum of 1% fluctuation from the average prediction errors for all sets of beam geometries reported in Figure 8.

Also, from the observation of prediction errors corresponding to each beam numbers in plan, Model I outperformed Model II for the mean dose errors on PTV, Body, Bladder, Rectum, Left Femoral Head and Right Femoral Head.

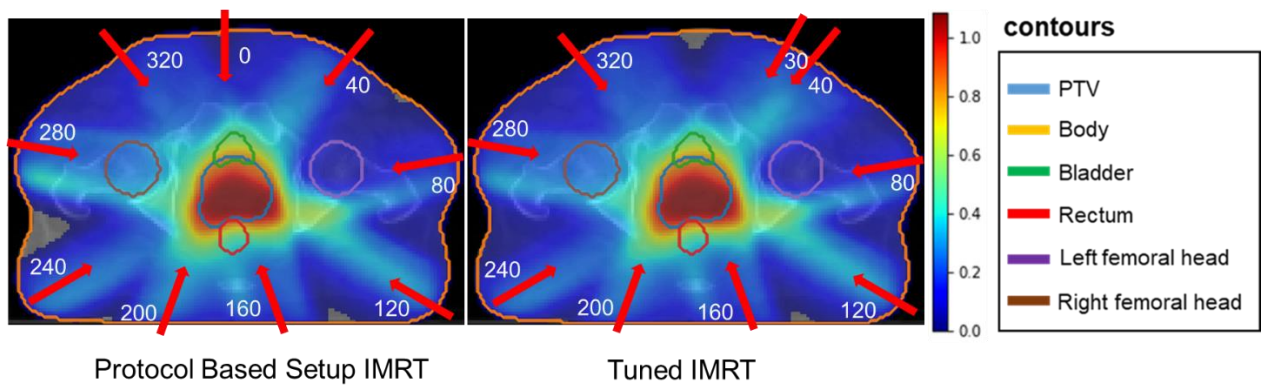

**Figure11:** Protocol Based Setup IMRT and Tuned IMRT dose distribution predicted by Models I.

Figure 11 shows the colored dose wash distributions in the PTV and in the organs at risk for Protocol Based Setup IMRT and Tuned IMRT. Protocol Based Setup 9 equidistant beam angles are: [0, 40, 80, 120, 160, 200, 240, 280, 320, 360] which are also shown in Figure 11 with red arrows.  By tuning one of the Protocol Based beam angles while keeping other parameters same as to that of Protocol Based Setup plan, we got the better plan than Protocol Based Setup IMRT plan. The Tuned IMRT beam angles with better plan are found to be [30, 40, 80, 120, 160, 200, 240, 280, 320, 360].  Visually, the dose distributions that are predicted from Models I for Tuned IMRT are similar to that of the Protocol Based Setup IMRT. Particularly, they are better matched in the high dose region surrounding the PTV, and have less matched in the lower dose regions.

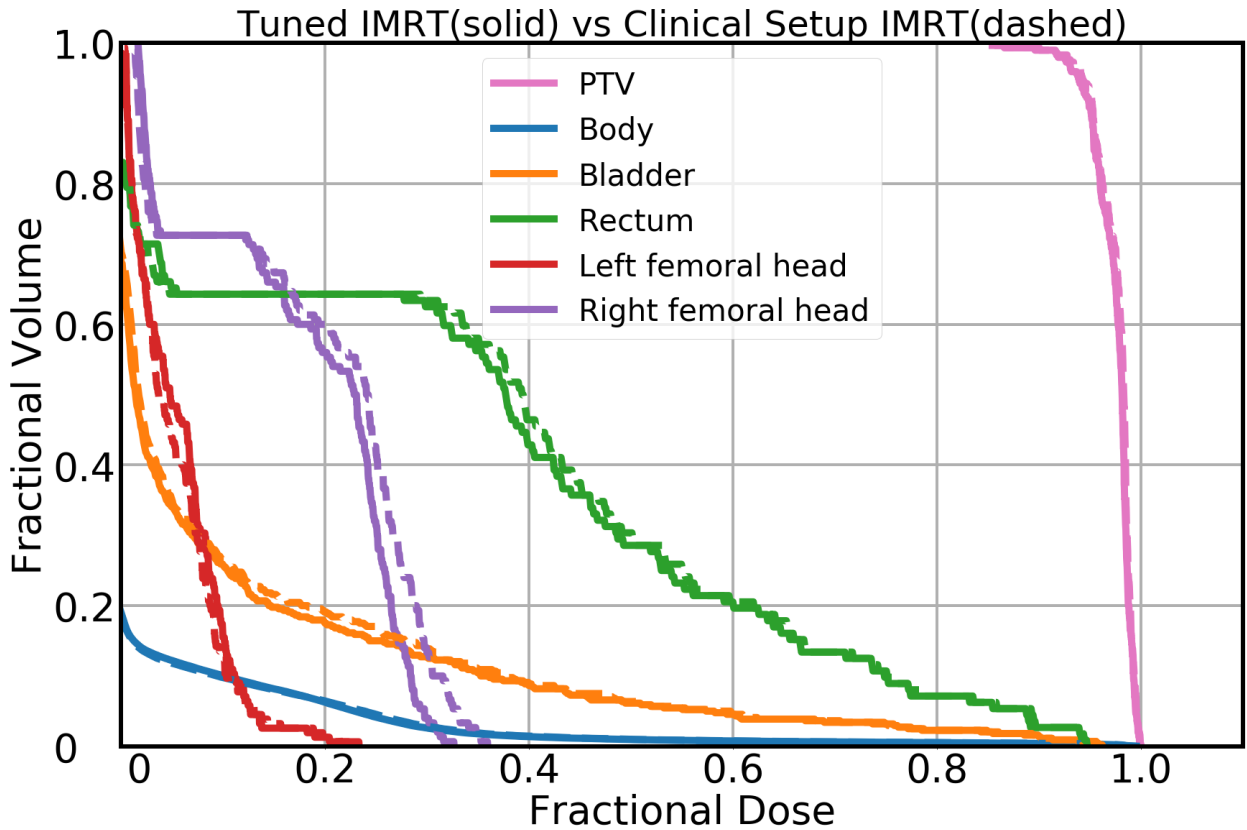

**Figure 12:** DVH plot obtained from Model I. The solid lines correspond to the Tuned IMRT, and the dashed lines correspond to the Protocol Based Setup IMRT.

DVHs obtained from dose predictions for the Protocol Based Setup IMRT plan and Tuned IMRT plan are presented in Figure 12. The Protocol Based Setup IMRT DVH curve corresponding to PTV are better matched with the Tuned IMRT DVH curve obtained corresponding to PTV. The DVH curves corresponding to other critical structures except without body show the fluctuations and the Rectum and the Right femoral head in the Tuned IMRT plan got less doses in comparison with the Protocol Based Setup IMRT plan. In addition Objective values calculated using Equations 2-3 for Protocol Based Setup IMRT and Tuned IMRT are 34.93 and 34.56 respectively. From these test cases, it is fully clear that Tuned IMRT plan is better plan than Protocol Based Setup IMRT plan.

## 4. Discussion:

The goal of this study was to predict Pareto optimal dose distributions by using anatomical structures and a varying number of beams (up to 10) and beam angles. To our knowledge, this is the first study to implement a deep learning-based, Pareto optimal dose prediction method with such flexibility of beam configuration. We designed deep learning models with two different types of input to represent the beam angles (Fig. 2). For Model I, we directly represented the beam angles as a binary vector for the second input. The anatomical structures used as the three-channel first input were planning treatment volume (PTV), body, and organs at risk (OARs). For Model II, we represented the beam angles as a conformal beam dose and included it as a fourth channel, with the 3 anatomical structure channels, in the model's single input. We generated conformal beam dose data that corresponded to the beam angles used in Model I (section 2.1).

Each model was trained, validated, and tested on 54, 6, and 10 patients, respectively. These patient data yielded a total of 27,000 training, 3,000 validation, and 5,000 testing plans. We used MSE as the loss function and the Adam optimizer for the model to minimize the loss between the ground truth dose and the predicted dose. The default learning rate of 0.01 for the optimization resulted in the best model for minimizing the validation loss. To avoid overfitting, we applied a dropout scheme in addition to group normalization, as shown in Table 3. Group normalization is more effective than batch normalization at handling small batch sizes.[67] Also, since group normalization is independent of batch sizes, using it helps to avoid manually selecting batch sizes for better convergence of the network.

Dose color washes in Figure 4 and the dose map differences in Figure 5 show that both models predicted dose distributions within the PTV more accurately than outside regions. This is because the PTV high-dose region is more uniform than other low-dose regions. Overall, our deep learning models (Model I and Model II) predicted voxel-level dose distributions that precisely matched the ground truth dose distributions.

The DVHs generated also precisely matched the ground truth (Fig. 6). Evaluation metrics—such as PTV statistics, dose conformity, dose spillage (R50) and homogeneity index—also confirmed the accuracy of PTV curves on the DVH (Table 4). PTV dose coverage error was within 4% for the prediction from Model I and within 9% for the prediction from Model II (Fig. 7).

Similarly, the predictions of mean, max dose over the PTV and the organs at risk and the mean of voxel wise difference over the body and 10% volume isodose reported in Table 5  and Table 6 respectively indicate the accuracy of both models. The average mean and max dose errors for the prediction from Model I were within 2% and 5%, respectively, for PTV, body, bladder, rectum, left femoral head and right femoral head. Likewise, the average mean and max dose errors for the prediction from Model II were within 6% and 12%, respectively, for PTV, body, bladder, rectum, left femoral head and right femoral head (Fig.8). Also, the mean voxel wise errors for the prediction from Model I and Model II were within 1 % for body and 10% volume isodose (Fig.9). All prediction errors reported in Tables 4, 5 and 6 represent the average errors for all 10 sets of beam geometries.

In addition, the prediction errors of mean dose corresponding to each beam numbers in each plan (1 to 10) over the PTV and the organs at risk, shown in Figure 10, are relatively even regardless of the number of beams the model was using to predict. The average mean dose errors for the prediction from Model I corresponding to each beam numbers in plan agreed with average errors for all 10 sets of beam geometries. The average mean dose errors for the prediction from Model II corresponding to each beam numbers were uniform with Body and Bladder in line with the average errors reported for all 10 sets of beam geometries. For PTV and the rest of the OARs' the prediction errors corresponding to each beam numbers were with maximum of 1% fluctuation from the average errors reported for all 10 sets of beam geometries.

 As we know that lower numbers of beam orientations generate low quality plans, and because low-dose region OARs such as femoral heads are further away from the PTV and have higher variability in the dose distribution, prediction errors are higher in these cases. The low prediction errors reported despite such variability in beam angles indicate that both models efficiently predict Pareto optimal dose distributions with high accuracy. However, Model I outperformed Model II in all evaluation criteria mentioned above.

For further verification, we performed t-tests to compare the prediction accuracy between the two models. In all cases, we found that p-values are extremely low (<0.001), which indicates that Model I's performance is statistically significantly superior to that of Model II. Model I outperforming Model II seems counterintuitive at first, as Model II's input of the conformal beam dose seems to be more similar to the final IMRT optimized dose that the model predicts. Although it is possible that including first-order priors or approximations as an input would improve a model's performance, a first-order approximation tends to differ from its exact version only on a local scale. For example, the difference between an accurate dose calculation engine, such as a Monte-Carlo–based engine, and an approximate one is the local scatter contribution from the primary beam. In our case, the difference between our conformal beam dose and the IMRT optimized dose is not local, because changing a beamlet's intensity affects the entire dose distribution along the beamlet's line through the body. This means that the Model II must then learn how to properly add and subtract values from the conformal beam, on a global scale, to transform it to the IMRT dose. This may be equally or more difficult for the neural network than attempting to generate the dose distribution from scratch, as it does with Model I. In addition, Model I is easy to implement and does not need to evaluate the conformal beam dose, which takes extra hours of work. Pareto optimal plans predicted for all 70 patients in this study give physicians the advantage of choosing among different tradeoffs for the critical structures. Physicians can quickly observe multiple Pareto optimal dose predictions in real time and ask the planner for further modifications to obtain the desired tradeoffs. Our method could be the clinical support tool that allows physicians to get the right treatment plan for each patient, and it would give the planner the advantage of making acceptable changes earlier, which would save time in treatment planning.

Dose color washes in Figure 11, DVH plots in Figure 12 and the objective values 34.93 and 34.56 calculated for Protocol Based Setup IMRT and Tuned IMRT justify the above statement.

A potential limitation of this study is the large sampling space from the beam number, beam angles, and the structure weights. There are currently 500 plans per patient, totaling to 35000 plans, and deep learning models do tend to interpolate well between data within its training distribution. However, the true number of needed samples to adequately cover the domain is unknown, and it is possible the model may fail on rare edge cases. Further investigation would be required to determine the number of samples needed in order to prevent such edge failures and to improve the model performance. An alternative could be to limit the sampling to only be in the clinically relevant space, which would drastically reduce the number of required samples.

Since Model I and Model II both predict well for prostate cancer, we plan to extend these models to other sites under the same IMRT setup. In addition, the Pareto optimal plans are not necessarily deliverable plans, since the machine parameters have yet to be calculated for the predicted dose. Since the additional sequencing steps after optimization may degrade the optimized plan, we plan to examine the extent of plan degradation the dose from applying a sequencing step, as well as investigate adding in the sequencing directly into the Pareto plan optimization as a direct aperture optimization. In addition, we plan to use a threshold-driven optimization engine called TORA,[73] to create deliverable plans from our current predicted doses in a high quality manner. In addition, we plan to extend this work to the Pareto optimal dose prediction for VMAT plans, given a tunable selection of arc orientations. We hope that this will help us to improve our automated treatment planning system.

While modern protocols for prostate typically ask for a set number of equidistant coplanar beam angles, this can sometimes be varied, where some beams can be dropped or added to tailor the plan to a specific patient. Therefore, we designed to keep the number of beam angles as a flexible parameter, from 1 to 10. In addition, as a future study, we wish to investigate the possible number of beam angles that can be reduced by having mathematically optimized beam angles. This can be achieved and studied by combining our present model with a deep-learning–based beam orientation optimization model.[21] This beam orientation optimization model can solve for a suitable set of beam angles, given a particular patient anatomy and structure weights. By combining these models, we will develop a framework that can provide plans that are tailored to each patient, in terms of both beam geometries and dosimetric criteria.

## 5. Conclusion:

We built U-Net–style deep learning models to predict Pareto optimal dose distributions of IMRT prostate plans involving anatomical structures and varying beam angles. We also compared dose predictions between two different beam configuration modalities. We found that both models efficiently predict Pareto optimal dose distributions with high accuracy. However, our deep learning model (Model I) that directly inputs the beam angles into the network as a binary vector was more accurate and robust than the state-of-the-art model (Model II) that inputs a conformal beam dose. Dose predictions from these models take less than a second, which would allow physicians to observe multiple predictions in real time and ask planners for further modifications to achieve the best tradeoffs. From this, planners will be able to make acceptable changes earlier, which will save time in treatment planning. We believe that our method of dose predictions will be a stepping stone to building automatic IMRT treatment planning.

In our future work, we plan to use a threshold-driven optimization engine to generate deliverable plans. By combining our dose prediction model and the beam orientation optimization model, we will develop a unified framework that can provide plans that are tailored to each patient, in terms of both beam geometries and dosimetric criteria. We will further concentrate on building sophisticated dose prediction and beam orientation optimization models that perform well in combination and that can also be independent of treatment sites.

## 6. Acknowledgements:

This study was supported by the National Institutes of Health (NIH) R01CA237269. The authors thank Dr. Jonathan Feinberg for editing the manuscript.